\documentclass[pre, showpacs, ,superscriptaddress]{revtex4-1} 
\usepackage{graphicx}
\usepackage{epstopdf}
\usepackage{dcolumn}
\usepackage{amsmath}
\usepackage{natbib}
\begin{document}

\title{Squeezing out the last 1 nanometer of water: A detailed nanomechanical study}
\author{Shah H. Khan}
\affiliation{Department of Physics, University of Peshawar, Peshawar, Pakistan}

\author{Peter M. Hoffmann}
\email{hoffmann@wayne.edu}
\affiliation{Department of Physics and Astronomy, Wayne State University, Detroit, MI 48201, USA}
\date{\today}

\begin{abstract}
In this study, we present a detailed analysis of the squeeze-out dynamics of nanoconfined water confined between two hydrophilic surfaces measured by small-amplitude dynamic atomic force microscopy (AFM). Explicitly considering the instantaneous tip-surface separation during squeezeout, we confirm the existence of an adsorbed molecular water layer on mica and at least two hydration layers. We also confirm the previous observation of a sharp transition in the viscoelastic response of the nanoconfined water as the compression rate is increased beyond a critical value (previously determined to be about 0.8 nm/s). We find that below the critical value, the tip passes smoothly through the molecular layers of the film, while above the critical speed, the tip encounters "pinning" at separations where the film is able to temporarily order. Pre-ordering of the film is accompanied by increased force fluctuations, which lead to increased damping preceding a peak in the film stiffness once ordering is completed. We analyze the data using both Kelvin-Voigt and Maxwell viscoelastic models. This provides a complementary picture of the viscoelastic response of the confined water film.
\end{abstract}
\maketitle

Nanoconfined liquids, which are of great interest in biology, geology, tribology and nanotechnology \cite{Fin96, Che12, Bhu95, Boc10, Dai10}, exhibit a variety of surprising nanomechanical and viscoelastic properties that are not yet well understood \cite{Kle98, Ant01, Dem01, Rav02, Bec03, Jef04, Pat06, Li07, Li08, Mat08, Kag08, Hof09, Kha10, Kha14}. Recently, a sharp transition was found as a function of compression rate in several liquids, where the liquid shows liquid-like characteristics at slow compression rate but solid-like peaks in relaxation times when compressed above a critical speed \cite{Zhu03, Pat06, Kha10}. This observation may depend on a critical lateral system size, for example the size of the confining tip in the case of atomic force microscopy (AFM) measurements \cite{Deb10, Deb11, Lab13}. The interpretation of this phenomenon depends on the conceptual model used to analyze the obtained data. Commonly used models are simple viscoelastic equivalent models \cite{Fin76}, especially the Kelvin-Voigt (KVM) and the Maxwell (MM) models, which are combinations of one purely elastic and one purely viscous element in either a parallel or a series configuration. These models roughly correspond to either a  solid-like (KVM) or a liquid-like (MM) system. In the usual interpretation of an AFM measurement, the elastic and viscous forces are additive, and thus correspond to a parallel or Kelvin-Voigt arrangement. However, in the study of liquids, a Maxwell model may be more appropriate, because it allows for stress relaxation. We have therefore used a conversion to a Maxwell model in previous work \cite{Jef04, Pat06}. Here, we present a detailed analysis of the nanomechanics of confined water, below and above the critical speed, using both models.

\section{Experimental}
A home-built small-amplitude dynamic AFM \cite{Pat05}  was used to record changes in the cantilever amplitude and phase. This AFM uses an optical fiber aligned to the back of the cantilever to measure changes in the amplitude with a resolution better than 0.1 \AA. Two cantilevers of stiffness 1.4 N/m and 2.4 N/m were used in the here presented measurements. The cantilevers were vibrated with amplitude less than 1 \AA  \ at off-resonance frequencies ($<$ 1 kHz), and the amplitude and phase were monitored using a lock-in amplifier.

Water was purified using a two-stage water purification system. In the first stage, reverse osmosis cleaned the water to an ion concentration of less than 10 ppm. This water was further polished using a Siemens (formerly USFilter) PURELAB classic UV/UF system. This system gives water with resistivity of 18.2 MΩ-cm (ion concentration below detection levels) with low organic contamination levels. Water from the same source was used for rinsing and cleaning the cantilever and the glass dishes. The Fisher brand micropipette was used to transfer water to the cell. The micropipette was soaked in concentrated HCL overnight before rinsing with clean water. The glass dishes were kept overnight in saturated NaOH isopropanol solution and washed in ultrapure water. 

The cantilevers were obtained from Nanosensors (PPP-FMR) and were used after dipping in piranha solution for one hour at about 100° C and then rinsing with deionized water. The cantilever stiffness was calibrated using both the geometrical method and the thermal method \cite{Mat06}. Freshly cleaved muscovite mica was used as substrate in a circular Kel-F liquid cell of radius 2.5 cm. The cantilever remained completely immersed in water during measurement. To minimize the effect of external vibrations and acoustic noise, the AFM was placed on a vibration isolation platform (Minus-K, BM4) inside a steel walled sound-proof box. This box was enclosed in another large box of the size of a phone booth. The cantilevers were vibrated with a frequency in the range from 400 Hz to 970 Hz. The range of frequencies used was restricted to avoid resonances of the systems, which would complicate data analysis. In this narrow frequency range, we detected no influence of the excitation frequency on the critical compression rate.

Unlike the surface force apparatus (SFA), AFM does not usually have a mechanism to measure the confinement size (tip-surface distance). In our case, we measured not only the amplitude and phase, but also the average photodiode signal, which corresponds to the deflection of the cantilever. The contact point with the surface was in most cases clearly discernable by a strong change in slope of the photodiode signal, a large reduction in amplitude and a significant phase change. Because of the very low apprach speeds we used, drift can be an issue.  This manifested itself by a non-zero slope of the photodiode signal far from the surface. In these cases, we applied a single linear drift correction to the entire measurement such that the resulting cantilever deflection was zero far from the surface and equal to the piezo motion when in full contact. With this method we were able to determine the tip -surface separation with an error of $\pm$ 0.2 \AA.

\section{Theory}
The simplest general viscoelastic models are the Kelvin-Voigt model (KVM) and the Maxwell (MM) model. In the KVM, the elastic and viscous elements are parallel to each other. Because the system relaxes back to its original shape after release of a stress on the system, the KVM is typically considered a model describing a solid. The MM, with its elastic and viscous elements in series, exhibits a permanent strain, as well as complete relaxation of any stress, and is therefore the simplest model of a liquid. In our case, we are dealing with a viscoelastic system that - on the time scale of our measurements - exhibits both liquid and solid characteristics. The standard analysis of AFM measurements treats the elastic and viscous forces of the sample as additive, and therefore assumes the KV model. This is because in AFM, the strain applied to the sample (change in tip-surface distance) is the same for the elastic and viscous "elements" of the sample, and the total stress (force) is measured, which is the sum of the elastic and viscous response forces. In this sense, the Kelvin stiffness and damping are the "real" measured values, which can be more directly related to other physical parameters, such as the viscosity \cite{Osh98, Kha14}, while the Maxwell model represents a mathematical transformation, which allows us to tease out additional features of the system response, as we will see below.

Both the KVM and the MM are associated with characteristic time scales. In the case of the KVM, the characteristic times is the retardation times, which is the time it takes for a strain to build up under an imposed stress. It is given by $t_{KV}=\gamma / k$, where $\gamma$ is the damping coefficient of the viscous element and $k$ is the modulus of the elastic element in the KVM. One would expect a perfectly elastic solid to have a vanishing retardation time, as the solid would deform instantaneously under an imposed stress. The characteristic time of the MM is the relaxation time, which corresponds to the time needed to relax an imposed stress on the system. A perfectly elastic solid would have infinite relaxation time, since plasticity or viscous flow is required to relax stress. The relaxation time is given by $t_{M} = \eta / R$, where $\eta$ is the damping coefficient and $R$ is the elastic modulus in the MM.
The characteristic times of both models can be easily converted into each other, according to:
\begin{equation}\label{converttimes}
t_{KV} = \frac{1}{\omega^2 t_M} 
\end{equation}
where $\omega$ is the angular frequency of the cantilever oscillation used to measure the mechanics of the system. It can be seen that the retardation and relaxation times are simple inverses of each other, scaled by the frequency of the dynamic measurement.
The corresponding elastic and viscous elements can also be converted into each other. For example, to convert from KVM elements to MM elements, we can use:
\begin{subequations}\label{convertelements}
\begin{align}
\eta &=\gamma+\frac{k^2}{\omega^2 \gamma} \label{convertelementsa} \\
R &= k+\frac{\gamma^2 \omega^2}{k} \label{convertelementsb}
\end{align}
\end{subequations}

\section{Results and Discussion}
\begin{figure}\centerline{\includegraphics[width=85mm, 
keepaspectratio]{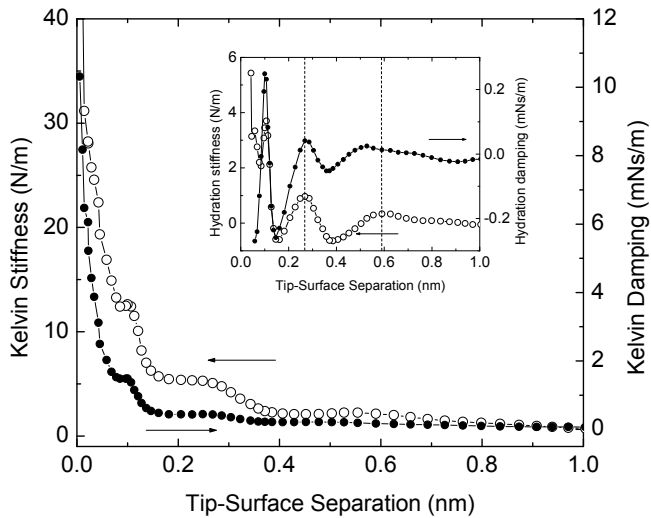}} \caption{Kelvin-Voigt stiffness and damping of nanoconfined water compressed at 2 \AA/s. Inset: Same data after subtraction of hydrophilic, repulsive background, showing effect of hydration only. Stiffness peaks are observed at 0.11, 0.27 and 0.59 nm. Stiffness and damping change smoothly and are in-phase.}\label{2AsKelvin}
\end{figure}

\begin{figure}\centerline{\includegraphics[width=85mm, 
keepaspectratio]{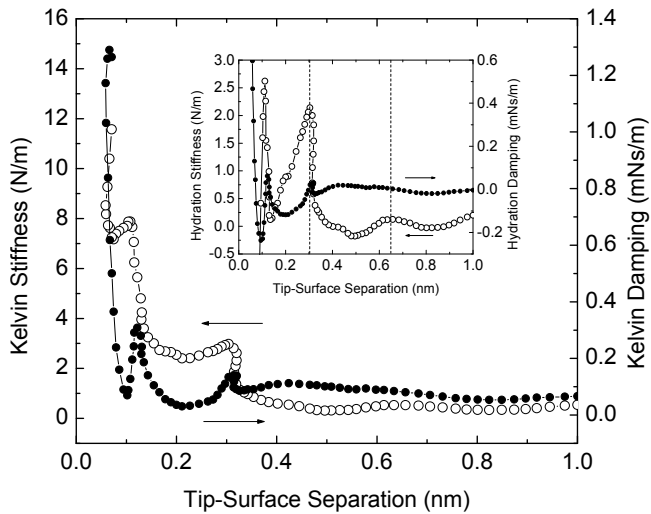}} \caption{Kelvin-Voigt stiffness and damping of nanoconfined water compressed at 14 \AA/s. Inset: Same data after subtraction of hydrophilic, repulsive background, showing effect of hydration only. Stiffness peaks are observed at 0.11, 0.30 and 0.64 nm. The tip motion shows pinning close to the location where the stiffness and damping exhibit peaks. The damping peak occurs during the sharp rise of the stiffness and therefore stiffness and damping are not in-phase.}\label{14AsKelvin}
\end{figure}

While in previous reports \cite{Jef04, Kha10}, we presented nanomechanical data as a function of piezo displacement, we here present  data plotted against actual confinement size. This was made possible by the development of a consistent measurement and drift correction method, as well as more thorough measurements. Figures \ref{2AsKelvin} and \ref{14AsKelvin} show KVM stiffness and damping data versus confinement size measured on nanconfined water with approach speeds of 2 \AA/s and 14 \AA/s, respectively. The insets show the same data after subtraction of the hydrophilic repulsive background.  Above a certain critical compression speed, as reported previously \cite{Pat06, Kha10} and seen by other groups \cite{Zhu03, Hof09}, we see a qualitative change in behavior, associated with the fact that above the critical speed the time to compress a single molecular layer exceeds a characteristic relaxation time of the system. This can be seen clearly by comparing Figures \ref{2AsKelvin} and \ref{14AsKelvin}: At slow compression (Figure  \ref{2AsKelvin}), the stiffness and damping are "in-phase" and change smoothly. By contrast, above the critical rate of about 8 \AA/s \cite{Kha10}, both damping and stiffness exhibit "pinning", i.\ e.\ the tip seems to rest on a liquid layer at certain tip-surface separations. At these separations, the stiffness rises sharply and the damping exhibits a peak (Figure \ref{14AsKelvin}). Then, as the pressure is increased, the stiffness reaches a maximum. After this, the tip "breaks through" a layer (a molecular layer is squeezed-out) \cite{Bec03} and the stiffness and damping reduce, before reaching a new pinning point. At this higher compression rate, the damping and stiffness are not in phase, because the damping peak is observed when the stiffness is still rising. This "out-of-phase" behaviopr of stiffness and damping at high compression rate is more clearly seen when the data is plotted versus piezo motion (which removes the compression of the data when plotted versus confinement size) \cite{Kha10}, but can also be seen here by close inspection. 

The fact that the damping (and therefore the effective viscosity) exhibits a peak at the pinning location suggests that as the load on an increasingly solid-like film is increased, the film structurally rearranges, leading to force fluctuations and higher dissipation. This is in agreement with simulations that showed that oscillatory damping in these systems is associated with structural changes in the layers\cite{Deb12}. Once a stable structure is formed, the stiffness of the system reaches a maximum and damping is reduced. Further increases in the pressure tilt the system towards disorder again, a molecular layer is expelled and the structured film gives rise to a more liquid-like, disordered film.

Stiffness peaks are observed at 2.9 $\pm$  0.2 \AA \ and 6.2 $\pm$ 0.3 \AA \ (as indicated by dashed lines in the inset figures), as approximately expected from the molecular size. However, suprisingly, plotting the data against actual tip-surface distance reveals that the first peak is located at about 1.1 $\pm$ 0.2 \AA \ from the surface. Comparing this finding with X-ray \cite{Che01} and simulation data \cite{Len06}, we find that this additional peak corresponds to {\em adsorbed} water molecules on the mica surface, while the subsequent peaks at $\approx$ 2.9 \AA \ and $\approx$ 6.2 \AA \ correspond to hydration layers. To our knowledge, this is the first time that AFM has been able to unambiguously resolve this features.

\begin{figure}\centerline{\includegraphics[width=85mm, 
keepaspectratio]{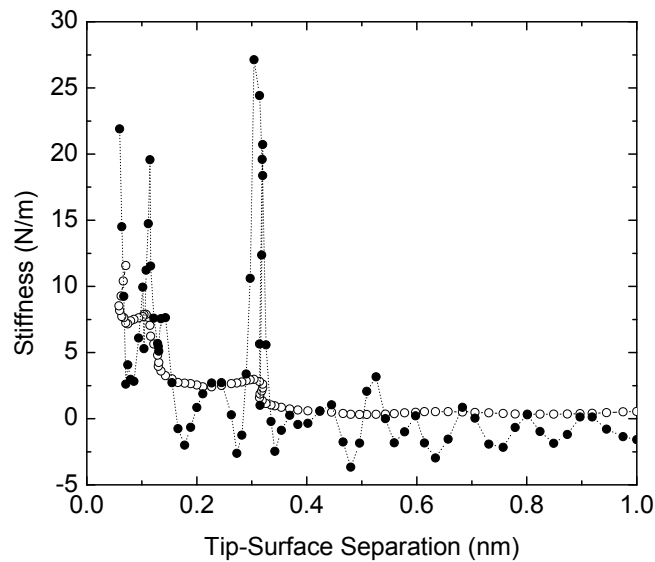}} \caption{Stiffness calculated from the cantilever bending (solid circles) compared to dynamically measured Kelvin stiffness (open circles).}\label{Stiffness}
\end{figure}

The observation that the tip is "pinned" at certain locations indicates a very high stiffness of the water layer. However, the dynamically measured stiffness in Figure \ref{14AsKelvin} shows only moderate peaks of about 3.0 N/m and 7.8 N/m, respectively. An alternative way to extract the stiffness of the layers is to take the derivative of the force calculated from the cantilever deflection, $z$, i.e. $k = k_L \rm{d} z/ \rm{d} (z+d)$. This data is very noisy and therefore needed to be average-filtered to clearly reveal patterns. The filtered stiffness data extracted from the cantilever deflection is shown in Figure \ref{Stiffness} and shows a sharp peak of almost 30 N/m at the first hydration layer and almost 20 N/m at the adsorbed layer. Why the difference between the stiffness obtained from the deflection and the dynamically measured stiffness? The most likely explanation is that the finite amplitude of the tip averages out  extremely sharp peaks in the stiffness, but brings the advantage of providing a smoother and less noise-prone signal.

\begin{figure}\centerline{\includegraphics[width=85mm, 
keepaspectratio]{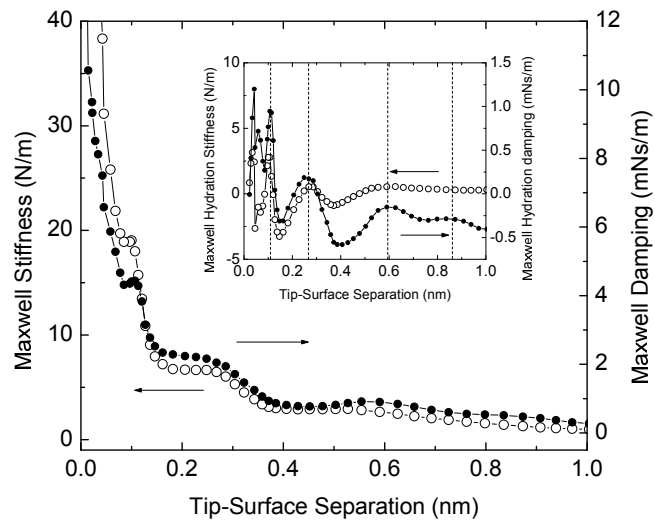}} \caption{Maxwell stiffness and damping of nanoconfined water compressed at 2 \AA/s. Inset: Same data after subtraction of hydrophilic, repulsive background, showing effect of hydration only. Stiffness peaks are at same locations as in Figure \ref{2AsKelvin}. A possible 4th peak is apparent at 0.86 nm in the damping data.}\label{2AsMax}
\end{figure}

\begin{figure}\centerline{\includegraphics[width=85mm, 
keepaspectratio]{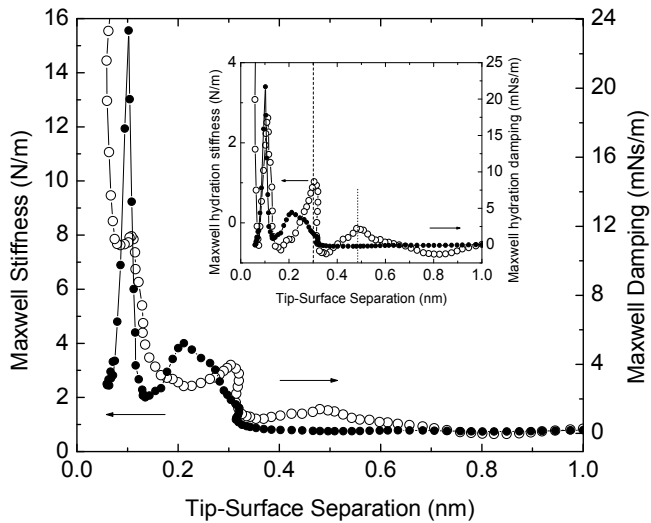}} \caption{Maxwell stiffness and damping of nanoconfined water compressed at 14 \AA/s. Inset: Same data after subtraction of hydrophilic, repulsive background, showing effect of hydration only. Stiffness peaks are at same locations as in Figure \ref{14AsKelvin}. The additional stiffness peak at 0.48 nm is not related to ordering, but comes from applying equation \eqref{convertelementsb} to a very low Kelvin stiffness at this location.}\label{14AsMax}
\end{figure}

Figures \ref{2AsMax} and \ref{14AsMax} show the data presented in Figures \ref{2AsKelvin} and \ref{14AsKelvin} converted into the Maxwell viscoelastic model. The MM should better correspond to a liquid system, as it shows viscous stress relaxation. Comparing the Kelvin and Maxwell picture for the 2 \AA /s data (Figs.\ \ref{2AsKelvin} \& \ref{2AsMax}), we see that the magnitudes of the stiffness and damping are increased when using the Maxwell model, while the location of the peaks and the overall shape of the data are quite similar for both viscoelastic models. The stiffness peak at 1.2 \AA \, which is associated with adsorbed water, has a magnitude of $k=$ 12.6 N/m in the Kelvin model and $R = $ 19.0 N/m in the Maxwell model. The difference between the damping values is more pronounced: Again for the adsorbed layer, damping in the Kelvin model, $\gamma$, peaks at 1.5 mNs/m, while in the Maxwell model the damping, $\eta$ is 4.4 mNs/m.  The ratio $\eta / \gamma = 1 +k^2/(\omega^2 \gamma^2)$ is a measure for how "elastic" the liquid responds. The fact that $\eta / \gamma$ is  larger than $R/k$ indicates that there is significant elasticity in the system.  However, this effect seems quite small, and the system response does not profoundly deviate from an overall liquid-like response. 

By contrast, in the 14 \AA /s data (Figs.\ \ref{14AsKelvin} \& \ref{14AsMax}), there are profound changes between the damping data in the Kelvin and Maxwell picture, while the stiffness is almost identical (at least close to the surface) in shape and magnitude. The Kelvin damping shows sharp peaks at the locations where the stiffness starts to rise and the tip is pinned on a molecular layer, and it decreases between the layers. By contrast, the Maxwell damping exhibits peaks between the layers.  Also, the Maxwell damping values are much higher than the Kelvin damping values by about an order of magnitude at idential locations (ignoring the additional peaks seen in the MM damping, which are even higher). Large Maxwell damping and low Kelvin damping indicates a more solid-like response, because in the Maxwell model, a large damping would indicate a very slow response of the viscous element to deformation, leaving the elastic element to take most of the strain under a quickly applied stress. In the Kelvin model, by contrast, with the two elements in parallel, a low damping allows the elastic element to take the stress quickly. 

\begin{figure}\centerline{\includegraphics[width=85mm, 
keepaspectratio]{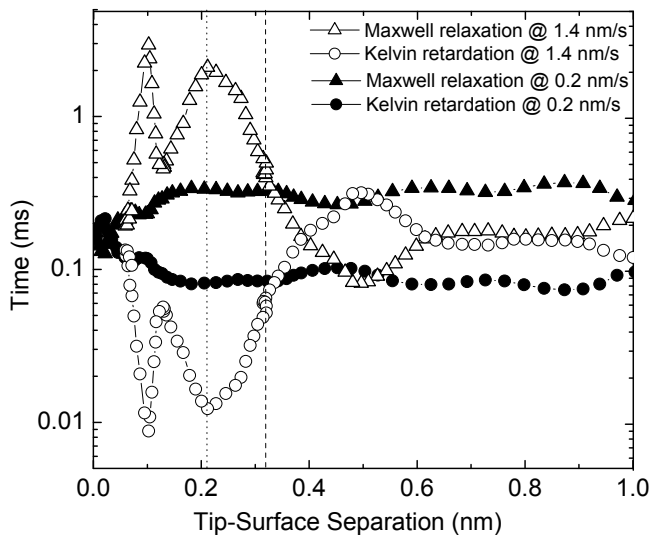}} \caption{Relaxation and retardation times}\label{Times}
\end{figure}

\begin{figure}\centerline{\includegraphics[width=85mm, 
keepaspectratio]{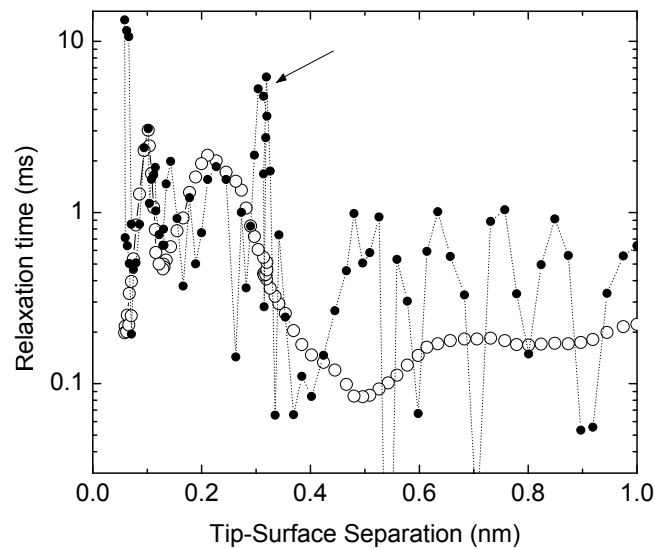}} \caption{Maxwell relaxation time calculated from the stiffness derived from the cantilever bending (solid circles) compared to the relaxation time calculated from the dynamically measured stiffness (open circles). Except for the different noise levels, the two relaxation time measurements agree quite well. The notable exception is the "solidification" peak in the relaxation time of the first hydration layer, indicated by the arrow.}\label{Relax}
\end{figure}

Figure \ref{Times} shows the Kelvin retardation and Maxwell relaxation times at the two approach speeds, calculated from the dynamically obtained stiffness and damping data. At 2 \AA /s, the retardation and relaxation times are essentially featureless (solid symbols), indicating uniform liquid-like relaxation behavior. By contrast, the 14 \AA /s data shows strong peaks and an exponential-like increase in the relaxation time (and a corresponding decrease in the retardation time). Surprisingly, the peak in the relaxation time does not correspond with the solid response (pinning) of the confined layer (which occurs at 1.2 and 2.9 \AA), but can rather be found in-between, at about 2.2 \AA. This relaxation time peak corresponds to the peak in the Maxwell damping seen in Figure \ref{14AsMax}, where we also note an offset between the damping peak and the pinning location. Although we did not previously notice this offset between the solid (pinning) response of the molecular layer and the peak in the relaxation time, it is also apparent in our previous measurements \cite{Kha10}. It is made more obvious here due to the data being plotted against actual tip-surface distance. The lack of a relaxation time peak at the pinning location in Figure \ref{Times} may be related to the "smoothing out" of the dynamic stiffness measurement, as mentioned above. Indeed, if we use the stiffness calculated from the cantilever deflection (Figure \ref{Stiffness}), we find a prominent peak at the pinning position, as shown in Figure \ref{Relax}. This peak corresponds to a "solid-like", elastic response of a stable, ordered film.

What then is the meaning of the in-between peak at 2.2 \AA? Mathematically, this peak in the Maxwell damping is the result of applying equation \eqref{convertelementsa} to a situation with low Kelvin damping and only a slightly reduced stiffness. Physically, this corresponds to a situation where the measured Kelvin damping is similar to the bulk-liquid value, something we have observed before \cite{Kha10}, while the stiffness remains relatively large because of the (negative) hydrophilic disjoining pressure. This background stiffness is unrelated (but additive) to the hydration-related stiffness oscillations. Thus at high compression speed, we have to distinguish between two solid-like responses, corresponding to different structural states of the film. At the pinning locations, the film structurally rearranges to find a low-energy ordered structure, which manifests itsef in a peak in the stiffness. The relaxation time peak pointed out with an arrow in Figure \ref{Relax} is due to the large increase in stiffness of this ordered state. As the pressure is further increased, order is destroyed and a molecular layer is squeezed out. The stiffness reduces to the background value. However, mechanically, the now disordered film also reacts solid-like - due to the still finite stiffness and great reduction in damping. The low damping suggests a low viscosity, liquid-like state, but because of the hydrophilic disjoining pressure the film elastically resists squeeze-out.

\section{Summary and Conclusion}

We have presented detailed studies of the squeeze-out dynamics of nanoconfined water layers below and above a critical squeeze-out rate. The picture that emerges from these studies matches well with previous experimental and computational studies. In particular, by using a direct measurement of the nanomechanics of the system, we confirm the existence of the adsorbed water layer on mica surface as suggestedby earlier studies \cite{Che01, Len06}. By analyzing the data using two different viscoelastic models, using improved data analysis methods to determine the true tip-surface separation, and determining the film stiffness in two distinct ways, we reconstructed the overall dynamics and viscoelastic response as a hydration layer is {\em rapidly} compressed  and expelled between two approaching hydrophilic surfaces. As seen in Figure \ref{14AsKelvin}, initially the stiffness is starting to rise and damping exhibits a peak, suggesting that the film is trying to obtain a stable structure under increasing load. The process of trying to obtain a stable structure leads to higher force fluctuations\cite{Deb12}, which result in higher effective damping. The stiffness continues to rise sharply and peaks at about 0.3 nm as a stable film is established that contains an integer number of layers (in this case, one adsorbed and one hydration layer). At this point, the damping starts to reduce as the film reacts more elastically and the structure is stabilized. As the film is further compressed, the stiffness drops, and the damping shaprply reduces to the bulk value. At this point the confined film is disordered and structurally liquid-like. The fact that, unlike the damping, the stiffness does not go back to bulk levels is due to the rapidly increasing hydrophilic compression pressure (or negative disjoining pressure). The finite stiffness gives rise to an elastic response, manifesting itself in an "in-between" peak in the Maxwell relaxation time. At about 0.15 nm, the same cycle repeats for the adsorbed layer. After this, the stiffness rises sharply as the tip encounters the solid surface. On the basis of our detailed analysis, we see that although both the Kelvin-Voigt model and the Maxwell model are equally valid, the Kelvin model seems to be more appropriate and straightforward in the interpretation of measurements obtained using experimental techniques like ours.

\begin{acknowledgments}
P.\ M.\ H.\ would like to acknowledge funding through NSF-DMR 0804283 and Wayne State University.
\end{acknowledgments}

\bibliography{KelvinMaxwell4}

\end{document}